\documentclass[fleqn,10pt]{wlscirep}
\usepackage[utf8]{inputenc}
\usepackage[T1]{fontenc}

\title{The Case for a Deep Search for Earth's Trojan Asteroids}

\author[1,*]{Renu Malhotra}
\affil[1]{Lunar and Planetary Laboratory, The University of Arizona, Tucson, AZ 85721, USA.}
\affil[*]{renu@lpl.arizona.edu}

\begin{abstract}
The existence of Earth's Trojan asteroids is not well constrained and represents a major gap in our inventory of small bodies in near-Earth space. Their discovery would be of high scientific and human interest.
\end{abstract}
\begin{document}

\flushbottom
\maketitle

\thispagestyle{empty}

\section*{}

It has been known for more than two centuries that minor planets can share the orbit of a planet in a dynamically stable state if they remain near the triangular Lagrange points, L4 and L5, leading or trailing a planet in its own orbit by $\sim$~60 degrees in longitude (Fig.~1).  Significant populations of such ``Trojan asteroids" are known for Jupiter, Mars, and Neptune.  However, only one Earth Trojan is currently known, 2010 TK7, discovered serendipitously by the WISE mission\cite{Connors:2011}.  This object, of absolute magnitude $H=20.6$ (diameter 300 m for assumed albedo of 0.1), has a large libration amplitude around L4 that brings it into close proximity to Earth every few hundred years; it was discovered near its closest approach distance of about 20 million kilometers. The orbital motion of 2010 TK7 is chaotic on megayear timescales, and the origin of this object remains undetermined at present~\cite{Connors:2011}.  The existence of a more stable and potentially primordial population of Earth Trojans (henceforth referred to as ETs) is thus not well constrained and represents a major gap in our inventory of small bodies in near-Earth space.

Surveys for near-Earth objects (NEOs) do not usually observe directly near Earth's L4 and L5 regions, owing to their unfavorable viewing geometry from Earth which requires observing during twilight or dawn hours when the background sky is bright. Moreover, 
because ETs are never near opposition in the night sky, they present only a partially sunlit hemisphere to Earth observers. Two decades ago, Whiteley \& Tholen [ref. \cite{Whiteley:1998}] bravely conducted a survey for Earth's Trojans from the University of Hawaii's 2.24-m telescope using an early generation of CCD detectors.  Although their search of the L4 region was frustrated by bad weather, they successfully surveyed 0.35 square degrees of sky near L5; they discovered no ETs brighter than the survey's limiting $R$ magnitude of 22.8. 

Last year, in February 2017, NASA's OSIRIS-REx %(Origins, Spectral Interpretation, Resource Identification, and Security Regolith Explorer) 
spacecraft conducted a survey for Earth's L4 Trojans when it cruised near the L4 point during its outbound journey to the asteroid Bennu; no ETs were discovered, and, based on the survey's sensitivity, an upper limit on the L4 ET population was estimated to be similar to that of the L5 ET population from Whiteley \& Tholen's twenty year old results~\cite{Cambioni:2018}.  Also last year, JAXA's Hayabusa-2 space mission surveyed for Earth's L5 Trojans when it cruised near the L5 point during its outbound journey to the near-Earth object Ryugu; no ETs were discovered, but implications of this null result for the upper limit to the L5 ET population have not been reported \cite{Yoshikawa:2018}. The current upper limits mean that the ET population could be as large as several hundred objects of a few hundred meters in size~\cite{Wiegert:2000}, and by extrapolation, many more at smaller sizes. 
For comparison, the population of all NEOs of diameter exceeding 300 m is estimated to be about ten thousand \cite{Harris:2015}. In other words, the unseen Earth Trojan population could be several percent of the entire NEO population, but confined to a much smaller volume near Earth's orbit.

Numerical simulation studies indicate that an ET population can orbit stably on timescales comparable to the age of the Earth~\cite{Tabachnik:2000,Malhotra:2011b}. While ETs in more eccentric and inclined orbits are strongly chaotic in Earth's orbital vicinity, those in low-eccentricity ($e < 0.1$), low inclination ($i <12^\circ$) orbits librating in proximity to L4 and L5 would be little perturbed. An ancient population in these zones would leak out only very slowly on gigayear timescales under the collective gravitational effects of planetary perturbations, although non-gravitational effects such as the Yarkovsky effect~\cite{Bottke:2006a} and collisions may affect these rates. 
 
The survival to the present day of an ancient ET population is reasonably assured provided Earth's orbit itself was not strongly perturbed since its formation.  It is therefore pertinent to consider that modern theoretical models of planet formation find strongly chaotic orbital evolution during the final stages of assembly of the terrestrial planets and the Earth-Moon system \cite{Morbidelli:2012}. Such chaotic evolution may at first sight appear unfavorable to the survival of a primordial population of ETs. However, during and after the chaotic assembly of the terrestrial planets, it is likely that a residual planetesimal population, of a few percent of Earth's mass, was present and helped to damp the orbital eccentricities and inclinations of the terrestrial planets to their observed low values, as well as to provide the so-called ``late veneer" of accreting planetesimals to account for the abundance patterns of the highly siderophile elements in Earth's mantle \cite{Raymond:2013}.  Such a residual planetesimal population would also naturally lead to a small fraction trapped in the Earth's Trojan zones as Earth's orbit circularized. 

In addition to potentially hosting an ancient, long-term stable population of asteroids, Earth's Trojan regions also provide transient traps for NEOs that originate from more distal reservoirs of small bodies in the solar system like the main asteroid belt. The only object currently designated an ET, 2010 TK7, with its large libration amplitude and its high orbital inclination, is itself a candidate for a temporarily captured NEO \cite{Connors:2011}. The temporary capture of NEOs into Earth-Trojan-like orbits is rather rare in computer simulations, suggesting a very small steady-state population of such objects of only about $16\pm3$ objects of diameter larger than 160 m [ref. \cite{Morais:2002}].  However, it is possible that unmodelled effects, such as the Yarkovsky effect \cite{Bottke:2006a}, not included in these simulations may enhance the temporary capture efficacy of the Earth's co-orbital region.  

Indirect evidence of a significant population of Earth co-orbitals may already be at hand in the well-preserved impact crater record on the Moon.  It has long been known that synchronously rotating moons in the solar system accumulate a higher density of impact craters on their leading hemispheres than on their trailing hemispheres~\cite{Zahnle:1998}. The ratio of crater density on the leading versus trailing hemispheres depends on the orbital distribution of the impactor population.  On the Moon, young craters (with ages $< 1$ Gyr) are about 70\% more frequent on the Moon's leading hemisphere than on its trailing hemisphere \cite{Morota:2005}, but the current NEO population models can account for only a $30\%$ higher frequency \cite{Ito:2010}.  
This discrepancy between the observations and the model-based predictions suggests the existence of an unmodeled population of impactors of low impact velocity,
 $v_{\rm impact}<10$~km~s$^{-1}$ (well below the $22$~km~s$^{-1}$ average impact velocity of NEOs), which would have a higher probability of impact on the leading lunar hemisphere \cite{Ito:2010}.   
The factor of $\sim$~2 discrepancy implies that there are twice as many low velocity impactors as are in the current NEO population model. 
Such an impactor population would have low eccentricity, low inclination, Earth-like orbits. Models of the origin of NEOs from the asteroid belt do not such objects in the numbers required to explain the discrepancy.  
However, primordial ETs, as they gradually leak out on gigayear timescales, are a natural candidate for the supply of such an unmodelled and presently unseen low-relative velocity population of impactors on the Moon and Earth.
   
A modern observational assessment of the ET population is important for a more complete inventory of the sources of meteoroidal impactors that pose a potential hazard to life on our home planet. Such an assessment would also provide novel constraints on the formation and dynamical history of Earth and of the inner solar system. If they exist, ancient Earth Trojans would have low relative velocity with Earth, making them attractive targets for exploration by spacecraft and even for human exploration.  Discovery and study of such objects would inform our understanding of the primordial building materials of our planet, including the provenance of its biocritical ingredients.
The complete absence at the present-day of ancient Earth Trojan asteroids should motivate a re-assessment of Earth's dynamical history and of the dynamical clearing of residual inner solar system planetesimals.  

To improve observational assessment of the ET population, astronomical surveys for faint NEOs with large telescopes could devote twilight and dawn time to search the Earth's L4,L5 regions. Follow-up of candidate discoveries would be critical to pin down their orbits so as to distinguish Trojans from non-Trojan NEOs. Ongoing programs such as the Catalina Sky Survey (https://catalina.lpl.arizona.edu/) and Pan-STARRs (https://panstarrs.stsci.edu/), the US National Science Foundation's forthcoming flagship project, the Large Synoptic Survey Telescope~\cite{LSSTColl:2009}, and the proposed NASA space mission NEOCAM~\cite{Mainzer:2016} hold promise to close this gap in our knowledge of asteroids in Earth's near-space neighborhood.

%\noindent LaTeX formats citations and references automatically using the bibliography records in your .bib file, which you can edit via the project menu. Use the cite command for an inline citation, e.g.  \cite{Hao:gidmaps:2014}.

%For data citations of datasets uploaded to e.g. \emph{figshare}, please use the \verb|howpublished| option in the bib entry to specify the platform and the link, as in the \verb|Hao:gidmaps:2014| example in the sample bibliography file.

\section*{Acknowledgements}

I thank NASA NExSS (grant NNH13ZDA017C) and NSF (grant AST-1312498) for research support.

%\section*{Author contributions statement}

%Must include all authors, identified by initials, for example:
%A.A. conceived the experiment(s),  A.A. and B.A. conducted the experiment(s), C.A. and D.A. analysed the results.  All authors reviewed the manuscript. 

%\section*{Additional information}

%To include, in this order: \textbf{Accession codes} (where applicable); \textbf{Competing interests} (mandatory statement). 

%The corresponding author is responsible for submitting a \href{http://www.nature.com/srep/policies/index.html#competing}{competing interests statement} on behalf of all authors of the paper. This statement must be included in the submitted article file.

\begin{figure}[ht]
\centering
\includegraphics[angle=35,width=\linewidth]{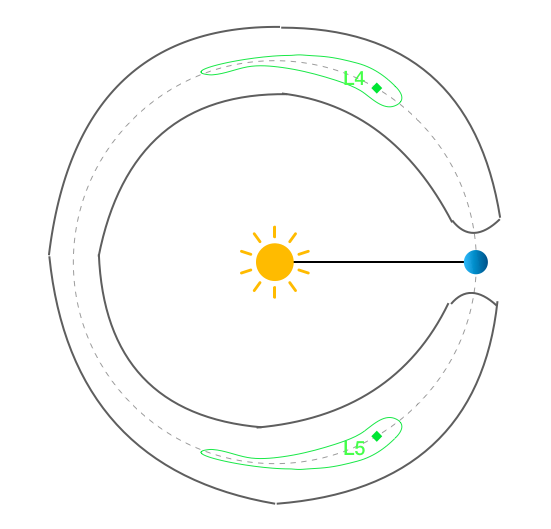}
\caption{A schematic diagram to illustrate the triangular Lagrange points (L4 and L5), Trojan asteroid orbits (in green) and horseshoe-type orbits (in gray). These are sketched in the rotating frame centered at the Sun and rotating with the planet's mean angular velocity.}
\label{fig:sketch}
\end{figure}

\end{document}